\documentclass[11pt,preprint]{aastex}
\usepackage{amsmath}

\usepackage{color}

\definecolor{darkgreen}{rgb}{0,0.4,0}

\def\AA{{\bf {A}}}
\def\BB{{\bf {B}}}
\def\JJ{{\bf {J}}}

\def\xx{{\bf {x}}}
\def\ee{{\bf {e}}}
\def\nn{{\bf {n}}}

\shorttitle{~}
\shortauthors{Pontin {\it et al.}}

\begin{document}

\title{Braided magnetic fields: equilibria, relaxation and heating}


\author{D.~I.~Pontin, S.~Candelaresi, A.~J.~B.~Russell and G.~Hornig}
\affil{University of Dundee, Nethergate, Dundee, DD1 4HN, UK}
\email{dipontin@dundee.ac.uk}

\begin{abstract}
We examine the dynamics of magnetic flux tubes containing non-trivial field line braiding (or linkage), 
using mathematical and computational modelling, in the context of testable predictions for the laboratory and their significance for solar coronal heating. We investigate the existence of braided force-free equilibria, and demonstrate that for a field anchored at perfectly-conducting plates, these equilibria exist and contain current sheets whose thickness scales inversely with the braid complexity -- as measured for example by the topological entropy. By contrast, for a periodic domain braided exact equilibria typically do not exist, while approximate equilibria contain thin current sheets. In the presence of resistivity, reconnection is triggered at the current sheets and a turbulent relaxation ensues. We finish by discussing the properties of the turbulent relaxation and the existence of constraints that may mean that the final state is not the linear force-free field predicted by Taylor's hypothesis.
\end{abstract}



\section{Introduction}
The magnetic field is a crucial driver of plasma dynamics in a wide range of environments. Many phenomena of interest involve explosive release of stored magnetic energy, mediated for example by ideal or resistive instabilities. Our knowledge of such instabilities tends to be based on rather ordered initial magnetic field configurations, in which -- for example --  large-scale twist or shear is present in the magnetic field. However, magnetic fields in laboratory and astrophysical environments are often inherently disordered, being characterised by field lines that are tangled with one another in non-trivial ways. Such tangled magnetic fields are the subject of this article: for our purposes we define a {\it braided magnetic field} as a magnetic flux tube in which the field lines have some non-trivial winding or linkage.

Braided magnetic fields have been used for some time to model  loops in the Sun's atmosphere, or {\it corona}, initially in response to Parker's proposed nanoflare heating mechanism \citep{parker1972,parker1988}. Therein, it is proposed that the corona is heated to the observed multi-million degree temperatures as a result of turbulent convective motions in the outer layers of the solar interior that tangle or braid the field lines about one another. This leads to an increase in magnetic energy in the coronal field that (it is proposed) is converted to kinetic/thermal energy by magnetic reconnection once the field becomes sufficiently complex. The efficiency of this heating mechanism remains a topic of heated debate -- the various modelling efforts were recently summarised by \cite{wilmotsmith2015}. While the predicted length scales of the field line braiding are below the spatial resolution of most instruments observing the Sun, recent direct observational evidence of braided loops in the corona has been claimed by \cite{cirtain2013}. Analogous processes will occur in any other astrophysical objects that comprise a turbulent, high-plasma-$\beta$ interior surrounded by a tenuous, low-$\beta$ corona (such as other stars and accretion disks).

{Field line tangling is important not only in astrophysical plasmas}, but in laboratory {and} fusion devices as well. In many spheromak or tokamak devices, ground state axisymmetric solutions exhibit regular structure with concentric flux surfaces. However, when this configuration is perturbed, as for example when the plasma is energised, regions of tangled, often ergodic field lines {are created}. Moreover, some tokamak configurations introduce stochasticity to the field near the plasma edge in an attempt to control the plasma properties there \cite[e.g.][]{jakubowski2006}. The ergodic field topology is crucial for transport within the device in question \citep{rechester1978}.

{In this paper we describe several important properties of braided magnetic fields, using some simple models of magnetic braids for illustration.}
In Section \ref{charsec} we introduce {useful measures that characterise magnetic braids.} In Sections \ref{idealsec} and \ref{nonidealsec} we investigate two aspects of the problem of energy release in braided magnetic {fields: the} existence of braided equilibria and turbulent resistive relaxation. We finish in Sections \ref{heatsec} and \ref{discusssec} with implications for coronal heating, and a discussion.

\section{Characterisation of braided magnetic fields}\label{charsec}
Braid theory deals with the tangling of a finite number of discrete strands. In this case, braids can be described using {\it braid words} that define the relative crossing of these strands \cite[e.g.][]{finn2007}. Discrete, strand-based models for braided coronal loops have also been developed \citep{berger2009}. However, magnetic fields are by nature space-filling, and here we focus on measures that treat the field and its linkage as continuous.
{The first set of tools identifies structures in a magnetic field from the field line mapping.} In magnetically open domains this constitutes a mapping from sections of the domain boundary on which {$\BB\cdot\nn<0$ (field lines enter the domain)} to those on which {$\BB\cdot\nn>0$ (field lines leave the domain)}. In solar physics, significant theory has been developed demonstrating that regions in which this mapping has large gradients are preferential locations for the formation of intense electric current layers \cite[][and references therein]{demoulin2006}. Such gradients are typically measured by the {\it squashing factor}
\begin{equation}\label{qeq}
Q=\frac{||{\rm D}F||^2}{\det({\rm D}F)}=\frac{\left(\frac{\partial X}{\partial x}\right)^2+\left(\frac{\partial X}{\partial y}\right)^2+\left(\frac{\partial Y}{\partial x}\right)^2+\left(\frac{\partial Y}{\partial y}\right)^2}{\left|\frac{\partial X}{\partial x}\frac{\partial Y}{\partial y}-\frac{\partial X}{\partial y}\frac{\partial Y}{\partial x}\right|},
\end{equation}
where $F=(X(x,y),Y(x,y))$ is the field line mapping from the `launch' boundary (coordinates $x$ and $y$) to the `target' boundary (coordinates $X$ and $Y$), and ${\rm D}F$ is its Jacobian. 
(For covariant expressions in general coordinates see \cite{titov2007}.) In periodic laboratory devices, on the other hand, the field line mapping may be iterated many times to produce Poincar{\' e} sections that reveal the presence of laminar and ergodic field regions \cite[e.g.][]{morrison2000}. Particular features of interest in this mapping for understanding the plasma dynamics include stable and unstable manifolds associated with fixed points \citep{borgogno2008,yeates2011b}.

{To quantify the  tangling in a magnetic braid, we transfer a result from fluid dynamics.}
Understanding the topology of the field line mapping is analogous to understanding mixing in a two-dimensional fluid that undergoes stirring.
Here the direction along the flux tube corresponds to the time in the stirring process, magnetic field lines to trajectories of fluid elements, and periodic orbits (field lines) can be interpreted as stirring rods. 
It is well established that there exist optimal {\it stirring protocols} for these stirring rods that yield the most efficient mixing of the fluid \citep{boyland2000}. A useful measure of the stirring quality is the {\it topological entropy}, {which} gives the exponential stretching rate of material lines in the fluid, for sufficiently complex mixing \citep{newhouse1993}. The faster such material lines grow in time, the more efficient the mixing. In the same way that one can calculate the topological entropy for a series of fluid particles trajectories in time, one can evaluate the entropy for field lines of a magnetic braid. Thus the topological entropy characterises how quickly initially adjacent field lines separate in an average sense.

{A final} important measure of magnetic topology is the {\it magnetic helicity}, {which quantifies the average linkage of field lines in the domain.} In a magnetically closed volume, $V$, {magnetic helicity is defined} by $H=\int_V\AA\cdot\BB \,{\rm d^3}x$, $\AA$ being the vector potential for the magnetic field $\BB$. When the volume of interest is not magnetically closed -- as in the case of a solar coronal loop where the magnetic field lines penetrate the solar surface -- the relevant quantity is the {\it relative helicity}
\begin{equation}
H_r=\int(\AA+\AA_{\rm p})\cdot(\BB-\BB_{\rm p})\,{\rm d^3}x.
\end{equation}
$H_r$ measures the helicity {(equivalently the average linkage)} relative to a reference field (usually the potential field) satisfying the same boundary conditions as $\BB$, where $\AA_{\rm p}$ is a vector potential for this {reference} field $\BB_{\rm p}$ \citep{berger1984}. Further information on the field line linkage may be obtained by considering the quantity 
\begin{equation}\label{curlya}
\mathcal{A}(\xx_0)=\int_{\ell(\xx_0)}\AA\cdot{\rm d}{\bf l}, \qquad \left.\nn\times\AA\right|_{\partial V}=\left.\nn\times\AA_{\rm p}\right|_{\partial V},
\end{equation}
where the integral is performed along a magnetic field line $\ell(\xx_0)$ passing through $\xx_0$ ($\partial V$ {and} $\nn$ being the boundary of $V$ and its outward normal, respectively). This {\it field line helicity} measures the net poloidal flux encircling the chosen field line \citep{berger1988}. It can also be shown to uniquely describe the magnetic topology of the field \citep{yeates2013} {and can therefore distinguish between different magnetic fields with the same total helicity}. The field line helicity, like the total and relative helicity measures, is an ideal invariant.

\section{Existence of braided equilibria}\label{idealsec}
\subsection{General considerations}
{The main purpose of} this paper is to investigate the importance of magnetic braiding for plasma dynamics, {and} the first issue we address is the existence of {smooth} stable equilibria for braided magnetic topologies. A key ingredient in Parker's magnetic braiding mechanism for coronal heating is the hypothesis that for sufficiently complex magnetic braids such equilibria do not exist. The coronal field is expected to evolve through a sequence of such equilibria, since the Alfv{\' e}n travel time along a coronal magnetic field line is much faster than the timescale of the photospheric driving. Plasma heating, mediated by reconnection, is then postulated when the field complexity reaches a state where no (continuous) equilibrium exists. Due to the low coronal plasma-$\beta$, such equilibria approximate Beltrami or force-free fields (f.f.f.):
\begin{equation}
\nabla\times\BB=\alpha\BB.
\end{equation}
Little has so far been unequivocally proven regarding the existence of such f.f.f.s in non-symmetric configurations. One exception is the study of \cite{bineau1972}, in which it was proved that in the absence of null points, f.f.f.s do exist  for sufficiently small $\alpha$ (though no bounds are provided on $\alpha$). There exist a number of different approaches to the problem, and -- while there is mounting evidence in favour of thin but finite current layers as opposed to the tangential discontinuities proposed by Parker -- there is presently no clear consensus, see \cite{craigsneyd2005,pontin2015a} and references therein. {We refer the reader here also to the work on equilibria that include flow and plasma pressure, and can be modelled by so-called double-Beltrami fields \citep[e.g.][]{yoshida2001}.}

\subsection{Model magnetic field}\label{modelsec}
\begin{figure}[!t]
\centering
\includegraphics[width=0.38\textwidth]{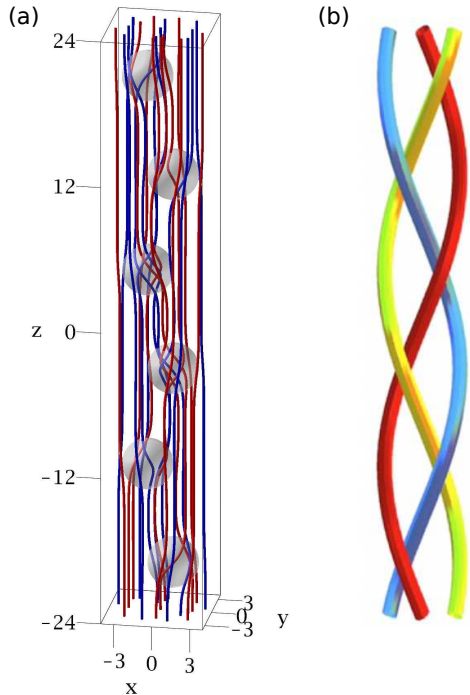}
\caption{(a) Selected field lines for the magnetic field defined by Eq.~(\ref{beq}) with $\kappa=1$; shaded isosurfaces of the magnetic field strength indicate the regions of magnetic-twist/current-density, which tangle the field lines. (b) Sketch of the `pigtail braid' structure formed by particular field lines.}
\label{blines}
\end{figure}
We consider in this paper a model braided magnetic field 
\begin{equation}
\BB=\sum_{i=1}^{6} \kappa (-1)^i \exp{\left( -\frac{(x-x_i)^2+y^2}{2}-\frac{(z-z_i)^2}{4}\right)}
\times (-y\,\ee_x+(x-x_i)\,\ee_y) ~+~1\ee_z \label{beq}
\end{equation}
where $x_i=(-1)^{i+1}$ and $z_{1..6}=\{-20,-12,-4,4,12,20\}$. When the parameter $\kappa=1$, particular  field lines in the domain (Figure \ref{blines}a) have a `pigtail braid' structure (Figure \ref{blines}b). {Such braiding could be generated in an initially uniform magnetic field via a sequence of rotational plasma motions on the boundary in a `blinking vortex' pattern \citep{aref1984}. However, the particular pattern of this braid is not crucial to the qualitative results presented. Rather, the salient feature is the generic property of tangling of magnetic field lines that is well established in laboratory plasmas and has recently begun to be quantified in solar coronal observations \citep{yeates2012,yeates2014}.}

\begin{figure}[!t]
\centering
\includegraphics[width=0.5\textwidth]{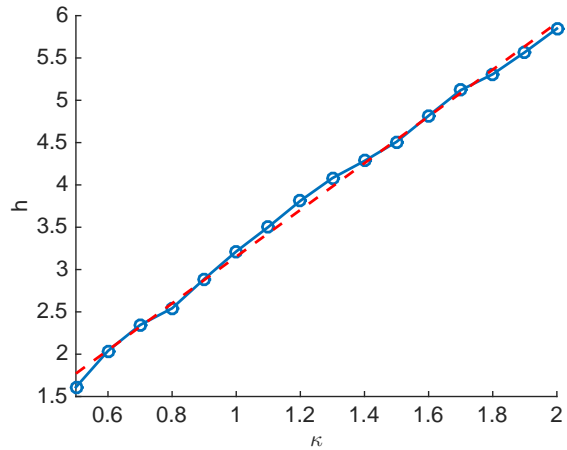}
\caption{Topological entropy $h$ as a function of the parameter $\kappa$ for the magnetic field (\ref{beq}), together with best linear fit (red).}
\label{h_vs_k}
\end{figure}

Here, we demonstrate for the first time that the topological entropy of this field  increases as the parameter $\kappa$ is increased, confirming that fields with higher $\kappa$ are ``more braided'' in a rigorous sense.  The calculation of topological entropy is performed by directly evaluating the growth of a material line under successive iterations of the field line mapping \citep{newhouse1993}, using an adaptive procedure in which additional points are included along the line in regions of high curvature, as required to resolve this length. The result is plotted for different values of $\kappa$ in Figure \ref{h_vs_k}. 
Fitting a straight line to the plot, we obtain a scaling
\begin{equation}\label{hscale}
h\sim \kappa\,(2.76\pm 0.09) \ + \ 0.39\pm 0.11
\end{equation}
(the errors being the 95\% confidence interval). That is, increasing $\kappa$ corresponds to increasing the overall field complexity as quantified by $h$. We return to consider this scaling later.

\subsection{Ideal relaxation simulations}
{This section} describes the results of a numerical relaxation method for investigating the existence of braided equilibria. The philosophy is to define a magnetic field with a given topology, and then relax towards an equilibrium while exactly preserving that topology. In order to guarantee topology preservation a Lagrangian numerical mesh is employed, as described in \cite{craig1986,candelaresi2014}. A magnetofrictional (MF) evolution enforces a monotonic decay of the magnetic energy in the system, as a  f.f.f.~is approached. The critical question is whether this equilibrium is smooth or contains tangential discontinuities, i.e.~current sheets. If it is discontinuous then the finite dissipation in a real plasma would lead to reconnection, topology change, and energy release. A number of studies using this approach have concluded that there is no evidence of formation of tangential discontinuities except in the presence of magnetic null points that we do not treat here \citep{craigsneyd2005,candelaresi2015}.
In particular the magnetic field of Equation (\ref{beq}) has been used as an initial condition for such ideal MF relaxations, first by \cite{wilmotsmith2009a} and  subsequently by \cite{candelaresi2015} using an improved numerical scheme. In neither case was any tendency towards the formation of tangential discontinuities observed. Of course no numerical relaxation can reach an exact equilibrium.
{Thus, here we describe} an extension of the MF relaxation. Specifically, we take the approximate equilibrium obtained after MF relaxation, and further relax this field using an MHD code  \citep{galsgaard1996} with a large viscosity applied. In this MHD code the resistivity is set explicitly to zero -- due to the sixth-order spatial differencing numerical dissipation (and thus topology change) is minimised, so long as length scales of variations in $\BB$ remain well above the grid-scale.

\begin{figure}[!t]
\centering
\includegraphics[width=5in]{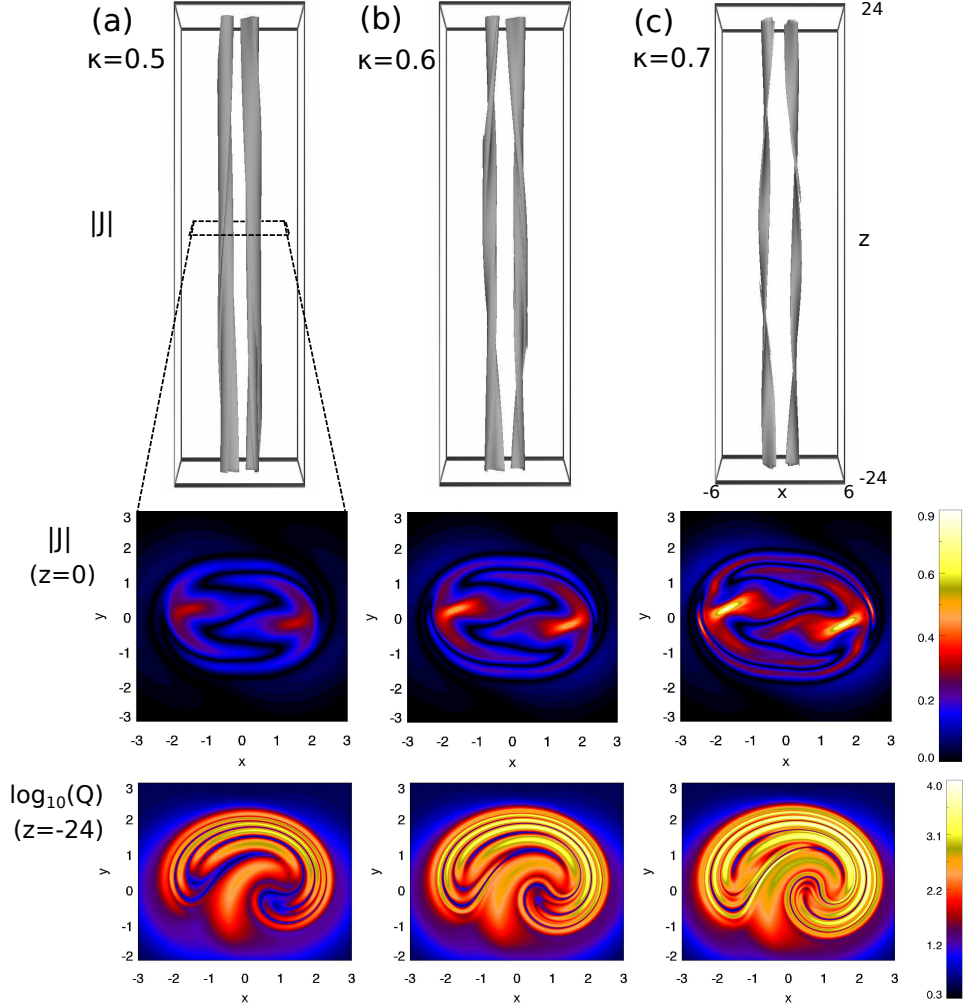}
\caption{Top row: $|{\bf J}|$ isosurface in the final state of the {\bf ideal} relaxation simulations for the magnetic field defined by Eq.~(\ref{beq}). Middle row: $|{\bf J}|$ in the midplane of the relaxed state. Bottom row: $\log_{10}(Q)$ plotted on $z=-24$. All for (a) $\kappa=0.5$, (b) $\kappa=0.6$, (c) $\kappa=0.7$.}
\label{jqideal}
\end{figure}

We have performed a systematic study varying the parameter $\kappa$ in Equation (\ref{beq}) using this approach, with both line-tied (perfectly conducting) and periodic boundary conditions in $z$ (for spatial domain $x,y\in [-6,6]$ and $z\in [-24,24]$). 
The results for the line-tied case were presented by \cite{pontin2015a}, who obtained a sequence of ideal equilibria for $\kappa<0.75$. The residual $\JJ\times\BB$ forces present in the approximate equilibria could be balanced by plasma pressure for a plasma-$\beta$ of order $10^{-3}$ -- thus the equilibria are as force-free as one would expect in the corona or a typical laboratory plasma. 
Figure \ref{jqideal} shows a sequence of the equilibria obtained, corresponding to different values of $\kappa$. The complexity of these magnetic braids increases for increasing $\kappa$, as shown by the topological entropy calculated above, which does not change during the ideal relaxation.
These equilibria -- that we emphasise are for the line-tied case -- contain thin but well-resolved current layers (Figure \ref{jqideal}, middle panels). The equilibria share the same topology as the magnetic field (\ref{beq}) for values of the parameter $\kappa$ up to $\kappa=0.75$. 

These results should be contrasted with the {newly considered} case where the $z$-boundaries are periodic. The simulations that we have performed using the same method but with periodic boundaries show no ideally accessible equilibrium for the range $0.5\leq\kappa\leq 0.75$. This non-existence of an equilibrium manifests through the generation of current layers that continue to thin until they reach the grid scale for all accessible numerical resolutions -- this leads naturally to reconnection via numerical dissipation. The reason for this absence of an equilibrium for the periodic magnetic braids is explained in the following section.

In summary, for periodic magnetic braids of the form in Equation (\ref{beq}), no equilibrium exists. For line-tied magnetic braids, a sequence of braided magnetic fields that are continuous equilibria does exist. However, these equilibria contain thin current layers, whose thickness is directly related to the field complexity, {becoming exponentially thinner as the parameter $\kappa$ is increased. We anticipate that this result for the line-tied case is a general one, not limited to the class of braids considered here, for the following reasons. First, the results are consistent with a number of other studies in which a magnetic field between two line-tied plates was subjected to sequences of shears, and an exponential thinning of current layers was observed  as the number of shears (and the resulting field complexity in the volume) increased \citep{vanballegooijen1988a,vanballegooijen1988b,mikic1989,candelaresi2015}.
Second, it can be argued that braided equilibria containing thin layers in the field line mapping {\it must} exhibit thin current layers on the same (or smaller) scale. This is explicitly demonstrated in the next section.

\subsection{Correlation length for the force-free parameter $\alpha$}	
\subsubsection{Exact equilibria}
{To further explore lengthscales in f.f.f.s}, let us consider the case of an exact f.f.f. $\nabla\times\BB=\alpha\BB$. {For} the periodic case any magnetic field containing ergodic field lines is consistent only with a linear f.f.f., or piecewise-linear in separate ergodic regions. For our model field (\ref{beq}), if we identify the top and bottom $z$-boundaries with one another to make the field periodic, 
then for $\kappa>0.5$ the region $\sqrt{x^2+y^2}\lesssim 2$ is highly mixed and Poincare maps show a volume-filling ergodic region. Thus, in this region $\alpha$ must be constant. Now, for a constant-$\alpha$ f.f.f.~$\JJ=\alpha\BB$ and the current helicity is $H_J=\int_V\JJ\cdot\BB \,{\rm d^3}x=\alpha\int_V |\BB|^2\,{\rm d^3}x$. In our case by symmetry we have $H_J=0$, which implies that $\alpha=0$. However, this is a contradiction, since it is inconsistent with the presence of ergodic field lines, and we conclude that an exact equilibrium for this field topology for the periodic case does not exist.

Consider now the requirements for an exact equilibrium in the case where $z=\pm z_0$ are line-tied, perfectly conducting boundaries. Then $\alpha$ is constant along field lines that run between these boundaries. To understand the nature of the distribution of $\alpha$ perpendicular to $\BB$, consider the following. Let us suppose that $\alpha$ has some global length scale, $\mathcal{O}(\ell)$, on some plane, say $z=-z_0$. Then, to find the length scales associated with the $\alpha$ distribution at, say, $z=+z_0$, we simply map $\alpha$ along field lines. We can relate the resulting length scales in the mapped quantity to the topological entropy, $h$, as follows. Recall that $h$ may be interpreted as the limit of the stretching rate of material lines under repeated application of the mapping, the lengths of these lines $L$ satisfing $L\sim\exp{(hn)}$ ($n$ being the iteration number). 

Consider a contour of $\alpha$ as such a material curve that is mapped in the `flow' of $\BB$. Then by definition the contour is stretched by a factor $\approx\exp(h)$ between $z=-z_0$ and $z=z_0$. Assuming constant area along the length of the braided flux tube, there must be a squeezing {on average} by a factor $\exp(-h)$ in the orthogonal direction. Thus, a typical contour of $\alpha$ on $z=+z_0$ will form an elongated structure with characteristic thickness $~L_\alpha\sim\ell\exp(-h)$. If the flux tube expands/contracts along its length then the area changes by a factor $B_n^{-}/B_n^+$, the ratio of the normal field components at the two boundaries for the elemental flux tube under consideration. Unless the braid has a high contraction/expansion factor this ratio will be $\mathcal{O}(1)$. Assuming an isotropic expansion, the length scales of $\alpha$ are modified to 
\begin{equation}\label{lalpha}
L_\alpha\sim\ell \, {\rm e}^{-h} \left(\frac{B_n^{+}}{B_n^-}\right)^{-1}.
\end{equation}
Now, in a f.f.f., $J_\|\propto\alpha$, and thus the equilibrium must exhibit current layers that are also on the length scale $L_\alpha$ (or smaller). Note that \cite{pontin2015a} made a similar argument based on the squashing factor $Q$ (see Eq.~\ref{qeq}), which shows that $L_\alpha\sim(Q\,B_n^+/B_n^-)^{-1/2}$.

\cite{pontin2015a} showed that the current layer thickness $w_J$ in the equilibria for line-tied boundaries follows the scaling
\begin{equation}\label{minwidtheq}
w_{J} =  \exp \left[  \kappa (-5.59 \pm 0.25)+2.53 \pm 0.18 \right].
\end{equation}
{Comparing this with the empirical scaling of $L_\alpha$ with $h$ that would be obtained from Eqs.~(\ref{hscale}) and (\ref{lalpha}),
we conclude that Eq.~(\ref{lalpha}) gives a rather weak upper bound on the current layer thickness  -- the current layers are in fact significantly thinner than the maximum permitted by the entropy argument. This difference} could be down to several factors. First, $h$ measures a global or average line stretching rate {(and the argument above treats a global-scale $\alpha$ contour), while the current layers form in the vicinity of selected field lines -- we assumed an equal squeezing of the $\alpha$ contour along its length to arrive at the factor $\exp(-h)$ while in practice this will vary between field lines. Second,} $h$ is determined by repeated iterations of the field line mapping, whereas in these line-tied simulations a material line or $\alpha$ contour is subject only to a single mapping iteration. The scaling of the current layer thickness in the domain can be shown to match very well with the scaling of the thinnest layers of $Q$ present in the mapping \citep{pontin2015a} -- this quantity being derived from a single iteration of the mapping and being minimised over field lines (over $x$ and $y$), thus removing the above two concerns.

\subsubsection{Approximate equilibria}\label{approxfff}
We addressed above the nature of an exact equilibrium -- however, any real (or simulated) magnetic field will only ever approximate such an exact f.f.f. Thus the question arises; what is the influence of a small departure from equilibrium on these results?
To answer this we define 
\begin{equation}\label{alphastarepsilonstar}
\alpha^* := \frac{{\bf J}\cdot {\bf B}}{B^2} = \frac{J_\|}{B}; \quad \epsilon^*:= \frac{\|{\bf J} \times {\bf B}\|}{\|{\bf J}\| \| {\bf B}\|} = \frac{\|{\bf J}_\perp\|}{\| {\bf J}\|}.
\end{equation}
The first quantity, $\alpha^*$, in general varies along field lines, but converges to the f.f.f.~parameter $\alpha$ as we approach a force-free equilibrium. The second quantity is a dimensionless function that measures the proximity to a f.f.f. We have $0 \le \epsilon^* \le 1$, where $\epsilon^*=0$ corresponds to an exact f.f.f.~and $\epsilon^*=1$ to a maximal non-force-free state. The above argument for the non-existence of force-free equilibria in a periodic domain relied on the fact that $\alpha$ is constant along field lines. However, a similar argument holds for a near-force-free state if $\alpha^*$ has only small variations over the length of the periodic domain, or in other words if the correlation length of $\alpha^*$, $l_{\rm corr}$ is longer than the domain length $L$. We define the correlation length for a given field line by
  \begin{equation}\label{correlationlength}
\frac{1}{l_{\rm corr}}:=    \left< \left|\frac{1}{\alpha^*}   \frac{\partial \alpha^*}{\partial s} \right| \right>,
\end{equation}
where  $s$ parameterises length along the field line and $\left<. \right>$ is an average over the field line. 
Following a similar calculation as in \cite{pontin2009} we use
\begin{equation}
0= \nabla \cdot {\bf J} = \nabla \cdot \left(J_\| {\bf e}_B \right) + \nabla \cdot {\bf J}_\perp = {\bf B} \cdot \nabla \alpha^* +  \nabla_\perp \cdot {\bf J}_\perp,
\end{equation}
which leads to 
\begin{equation}  
\left|\frac{1}{\alpha^*}  \frac{\partial \alpha^*}{\partial s} \right|= \left| \frac{1}{ J_\|/B} \frac{{\bf B}\cdot \nabla \alpha^*}{B}\right|=  \left|\frac{-\nabla_\perp \cdot {\bf J}_\perp}{J_\|} \right|\approx \frac{\epsilon^*}{d},
\end{equation}
where we used in the last step that for approximately force-free states $|J_\| | \approx \| {\bf J}\|$ and introduced a typical length scale $d$ for variations perpendicular to the field line: $\|{\bf J}_\perp\|/d \approx |-\nabla_\perp \cdot {\bf J_\perp} |$.  The result, $l_{\rm corr} \approx \left<d /\epsilon^* \right> $ shows that even for high-quality force-free approximations with $\epsilon^* \approx 10^{-3}$ (as  in our examples) the correlation length can be comparable with or smaller than the length of the domain (48 units in our case) if the perpendicular length scales become small as well ($d \approx 0.05$ for $k=1$ in Eq.~\ref{minwidtheq}). One can turn this argument around to make a statement about the structure of approximate equilibria with ergodic field lines.
Specifically, an approximate f.f.f.~with ergodic field lines in a periodic domain of length $L$ must have  $l_{\rm corr}\leq L$. This allows us to place an upper bound on the smallest perpendicular length scale for field lines in the ergodic domain $d_{\rm min} \le \left< d/\epsilon^* \right> \left< \epsilon^*\right> \le L \left< \epsilon^*\right>$, hence the field must contain  current sheets of thickness  $L  \left<\epsilon^*\right>$ or smaller. We note that this argument makes concrete the assertions of \cite{taylor1993} regarding filamentation of the current in the relaxed state of a tokamak plasma. Note further that \cite{candelaresi2015} have examined the correlation length in numerical relaxation simulations, and found that even when it becomes large on average,  it may remain relatively small at areas of high complexity, hindering the numerical relaxation process.

\section{Turbulent relaxation}\label{nonidealsec}
\subsection{Properties of the relaxation process}\label{turbprop}
The results described in the above section demonstrate that any sufficiently complex magnetic braid will be unstable in a resistive plasma owing to the development of thin current layers in the (approximate) equilibrium. Thus, in this section we describe the evolution following the onset of reconnection. This evolution was first followed in resistive MHD simulations with line-tied boundaries, starting from the magnetic field (\ref{beq}) (with $\kappa=1$), by \cite{wilmotsmith2010,pontin2011a}. It has subsequently been shown that repeating these simulations with periodic boundaries in the $z$-direction does not effect the qualitative nature of the relaxation although periodic boundaries do allow the field to reach a slightly lower energy end state \citep{yeates2015}.
In this section we make a spectral analysis that demonstrates the turbulent properties of the relaxation for the first time and we briefly summarise the main features of the evolution and final state.
 {The analysis that we present builds on the simulation of \cite{pontin2011a}, 
as well as two new simulations with the same setup and method, but with $\kappa=0.7$ and $\kappa=1.5$}, {thereby  exploring for the first time the impact of the braid complexity.}
{For concreteness we describe results of simulations with line-tied boundaries -- however, the general results apply equally well to the periodic case.}

\begin{figure}[!t]
\centering
\includegraphics[width=\textwidth]{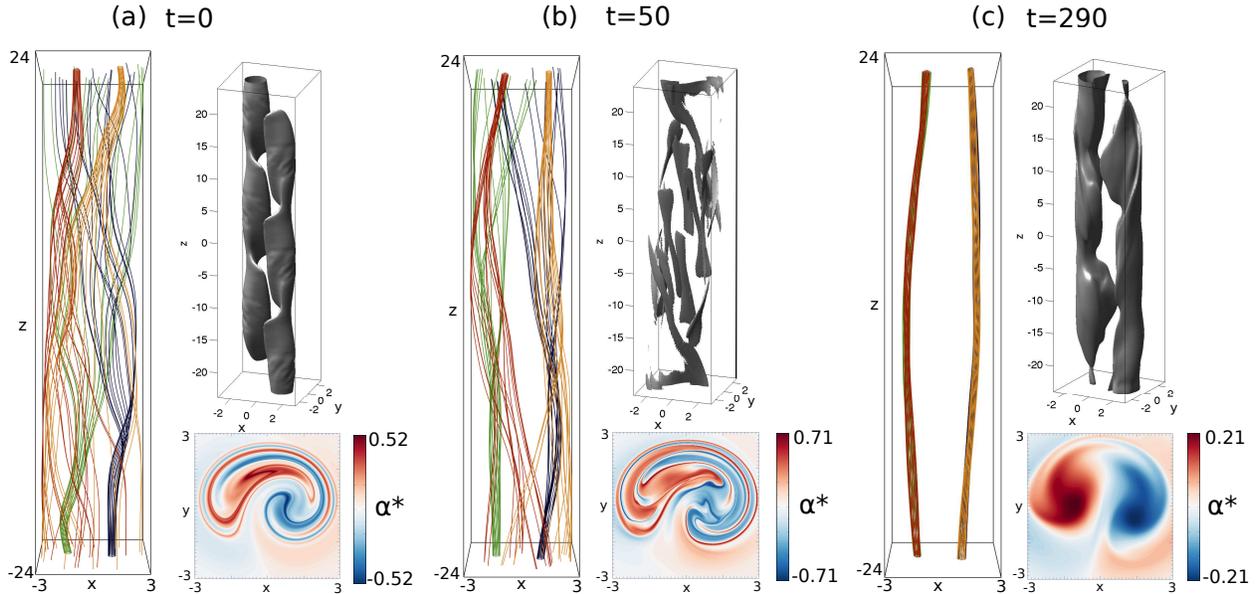}
\caption{Selected field lines (traced from fixed boundary footpoints), current isosurfaces, and contour maps of $\alpha^*$ on $z=-24$; each for the resistive relaxation simulation with $\kappa=1$.}
\label{res_composite}
\end{figure}
The resistive relaxation of our magnetic braid involves initially the formation of two current layers (those that form under the ideal relaxation), that due the finite resistivity undergo reconnection, leading to a cascade of further reconnection events. The global evolution of the system is characterised by an array of current layers that form and then dissipate, efficiently filling the volume of braided field lines -- see Figure \ref{res_composite}(b). The dominant field component in the current layers is the `guide' field along the tube (in the $z$-direction). Therefore the reconnection in these current layers takes the form of a continuous `flipping' or `slippage' of field lines \citep{priest1995,hornig2003}. The net result of these many localised reconnection events (around 30 when $\kappa=1$ and the magnetic Reynolds number $R_{\rm m}=10^{-3}$) is that the magnetic field `unbraids', leading to a much simpler topological state (Figure \ref{res_composite}c).

\begin{figure}[!t]
\centering
\includegraphics[width=0.5\textwidth]{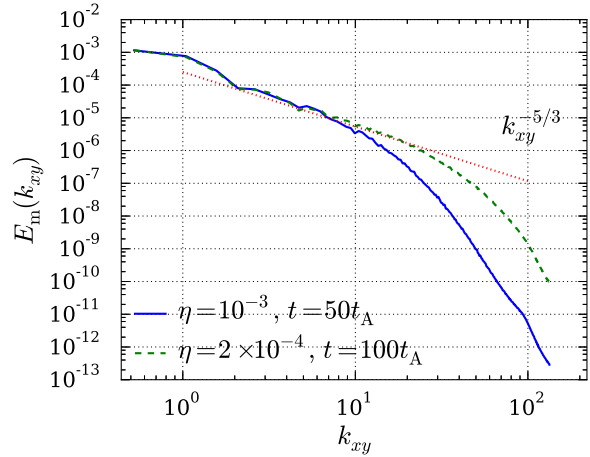}
\caption{2D spectra (averaged over $z$) for two resistive MHD relaxations of the field (\ref{beq}) with $\kappa=1$, with resistivity $\eta$ as indicated. Plotted are the shallowest spectra observed over time in each case (maximally turbulent state).}
\label{spectra}
\end{figure}
The resistive relaxation process described above has many of the properties of decaying turbulence. In Figure \ref{spectra} we plot the 2D magnetic energy spectrum over the $xy$-plane (averaged over $z$) for two of the simulation runs of \cite{pontin2011a}. We have {an estimated spectral index close to $-5/3$}, which is within the range expected for decaying 2D MHD turbulence \citep{biskamp2001}. We also see that as $R_{\rm m}$ is increased we obtain a larger inertial range over which this scaling is obeyed, as expected. {Together with this magnetic energy spectrum, a similar spectrum of kinetic energy is observed. The small-scale flows generated during the turbulent phase organise into decaying large scale flows in the final quasi-equilibrium (described below). This is consistent with the ``reverse dynamo" flow generation mechanism of \cite{mahajan2005}.}

A number of factors control how fully developed the {turbulence} becomes -- one being the value of $R_{\rm m}$ as shown in Figure \ref{spectra}. In addition, analysing the simulations with different  values of $\kappa$ (the initial field complexity), we find that the turbulence becomes more developed and the overall relaxation timescale increases as $\kappa$  is increased. This is illustrated in Figure \ref{jiso_kres} by isosurfaces of the current density for two resistive relaxation simulations with different values of $\kappa$. The plots are made at a time when a maximal number of discrete current layers appear in the volume -- at $t=55$ for $\kappa=1.5$, and $t=45$ for $\kappa=0.7$. Note for reference that the Alfv{\' e}n travel time along the length of the simulation domain in $z$ is approximately 48 time units. We observe that the number of small scale current layers is greater for greater $\kappa$. The time taken to reach the final relaxed state (the nature of which is discussed in the next section) is estimated to be $t=120$ for $\kappa=0.7$, $t=200$ for $\kappa=1$, and $t\gtrsim 400$ for $\kappa=1.5$ (or approximately 2.5, 4 and 8 crossing times, respectively).
\begin{figure}[!t]
\centering
\includegraphics[width=0.5\textwidth]{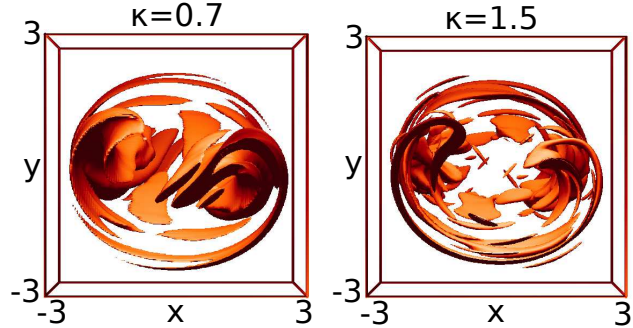}
\caption{$|{\bf J}|$ isosurface (at level 50\%) in the {\bf resistive} relaxation at a time when the field is maximally turbulent, for  $\kappa=0.7$ (left) and $\kappa=1.5$ (right), at times $t=45, t=55$, respectively.}
\label{jiso_kres}
\end{figure}

\subsection{Nature of the final state and constraints on relaxation}	
Taylor's relaxation hypothesis posits that the only conserved {topological} quantity during any sufficiently turbulent relaxation is the {total} magnetic helicity, and a variational calculation shows that the final state should be the linear force-free field \citep{Woltjer-1958-489-91-PNAS, taylor1974}. The theory was very successful in predicting the relaxed state  in a reverse-field pinch and to a lesser extent in other devices \citep{ortolani1993}, although more recently some departures of the final equilibrium from the `Taylor state' have been noted experimentally \citep{capello2008}. It has also been proposed that the hypothesis may apply in the Sun's corona \citep{heyvaerts1984}. 

A major result of the study of \cite{pontin2011a} was that the Taylor (linear f.f.f.) state is not reached after the turbulent relaxation of the magnetic braid (\ref{beq}). Since the net magnetic helicity  is zero, Taylor's hypothesis would predict the uniform field $\BB=1\ee_z$ as the final state. However, what is found instead is a non-linear f.f.f.~containing a pair of oppositely-twisted flux tubes, embedded within the uniform background field. This is shown by the plots of $\alpha^*$ (see Eq.~\ref{alphastarepsilonstar}) in Figure \ref{res_composite} 
 (recall that {for an exact f.f.f.,} this coincides with the force-free parameter $\alpha$).
The fact that the equilibrium is a {\it non-linear} f.f.f.~implies the presence of additional constraints on the relaxation.

 {One such constraint, identified by \citet{yeates2010}, is the {\it topological degree} \citep{polymilis2003} of the field.} This quantity is preserved during the relaxation so long as the topological degree of the boundary remains unchanged (as for example when the turbulent relaxation is confined to the interior) {and it may render the Taylor state inaccessible}. For the magnetic braid of equation (\ref{beq}) this topological degree -- given by the sum of the indices of periodic orbits of the field -- is 2, which is inconsistent with the predicted Taylor state. 
{A further promising tool is the {\it field line helicity}, defined as in Eq.~(\ref{curlya}). 
Unlike $H$, its distribution {\it uniquely} characterises the topology of the magnetic field \citep{yeates2013}, 
so understanding its evolution during turbulent relaxation 
could in principle allow prediction of the topology of the final state. 
Recently, \cite{russell2015} derived an evolution equation for the field line helicity, and analysed the importance of 
its different terms for slipping reconnection in a complex magnetic field. 
The reconnection primarily rearranges line helicity while having a relatively small impact on the total helicity, providing a new justification that relaxation approximately conserves total helicity while changing how the helicity density is distributed.  However, the relaxation cannot exchange field line helicity arbitrarily, 
which means the Taylor state may be prohibited in some cases, even when the topological degree would permit it. This could potentially explain why the contour map of $\alpha^*$ shown in Figure \ref{res_composite}(c) is not piecewise-constant, contrary to what would be expected if locally Taylor states (linear f.f.f.s) were produced in each of the flux tubes.}

\section{Implications for coronal heating}\label{heatsec}

We return now to the implications of the braiding mechanism for heating the solar corona.
We argued in Section \ref{idealsec} that in (either exact or approximate) braided equilibria, thin current layers must be present. Moreover, these current layers become increasingly thin and intense as the braid complexity -- as measured for example by the topological entropy -- increases. In practical terms, the implication is that if one considers a line-tied magnetic field that is driven slowly (compared to the Alfv{\' e}n speed) towards a state of high topological entropy, then onset of reconnection and energy release is inevitable. This is because as the field complexity increases  the corresponding current sheets in the volume get progressively thinner until a threshold for reconnection onset, determined by the plasma parameters, is reached. 

Thus energy may be stored in the magnetic field by the braiding mechanism, and the degree of braiding provides an onset threshold for energy release, such a threshold being required based on energy balance arguments \citep{parker1988}. The previous best theory to explain this threshold involves a monolithic current layer and requires the presence of anomalous resistivity \citep[e.g.][]{dahlburg2005}, while the critical degree of field line braiding that we propose here invokes the true complexity of the coronal field. The expectation, then, is that the coronal field will exist in a marginally-stable state in the vicinity of this critical degree of braiding, in which the boundary driving (which on average increases complexity) is balanced by the reconnection and energy release (which decreases complexity). One crucial measurement required for determining the heating efficiency of the mechanism is the free energy present in the field when the onset threshold for energy release is reached. For the magnetic field of Equation (\ref{beq}), this was estimated (assuming characteristic loop length of 50~Mm and field strength of 10-100~G) by \cite{pontin2015a} to be in the range $10^{25}-10^{28}$~ergs. This is more than sufficient for typical nanoflare models. However, there remain significant open questions to be addressed in assessing the contribution of the braiding mechanism for heating the corona. In particular, estimation of the timescale on which the braid complexity increases in the corona is non-trivial; some initial studies have been performed \citep{yeates2012,yeates2014}, but further work is required. In addition the timescale for energy release and the associated plasma response require further study.

\section{Discussion}\label{discusssec}
There is significant observational evidence that astrophysical plasmas are generically turbulent \citep[see][and references therein]{BrandenbSubramanianReview2005, lazarian2012,cranmer2015}. Therefore, the associated magnetic fields are characterised by a chaotic field line wandering, in contrast to the laminar, `combed' magnetic field lines of typical models. While laboratory plasmas are in general initiated with symmetric, well-controlled initial conditions, the onset of instabilities may quickly lead to the generation of ergodic field regions. 

In recent years there have been significant advances in modelling of complex, braided magnetic field structures. These advances have led to an understanding that the global field complexity plays a critical role in the plasma dynamics. 
{To date there have been few attempts in the laboratory to characterise the nature of field line braiding.  One relevant set of experiments has been run in recent years on the Large Plasma Device at UCLA. These experiments involve reconnection between flux ropes that are line-tied at one end, the reconnection being shown to occur in layers of high $Q$ between the ropes \citep{lawrence2009,gekelman2012}. The magnetic field structure that results from the reconnection has been studied in detail by
\cite{gekelman2014}, who measured the generation of chaotic field in the presence of three interacting, reconnecting magnetic flux ropes. They identified the region of highest field complexity as the vicinity of the reconnection site -- this being consistent with recent results on the topology resulting from current layer instabilities during 3D reconnection \citep{daughton2011,wyper2014b}. While measurements in such a plasma are extremely challenging, it is a promising future avenue to investigate whether braided/chaotic fields can be generated in this way, how the degree of braiding can be controlled, and the subsequent implications for the plasma dynamics.}
%

The modelling advances described in this paper have generated a number of testable predictions that are ripe to be studied in the next generation of laboratory plasma experiments.  The main results can be summarised as follows:
\begin{enumerate}
\item
In line-tied magnetic geometries, smooth braided equilibria do exist. However, these equilibria must contain current layers whose thickness becomes increasingly small for increasing field complexity. In practical terms, the implication is that if one considers a line-tied magnetic field that is driven slowly (compared to the Alfv{\' e}n speed) towards a state of high topological entropy, then onset of reconnection and energy release is inevitable. This is because as the field complexity increases  the corresponding current sheets in the volume get progressively thinner until a threshold for reconnection onset, determined by the plasma parameters, is reached. This result has particular importance in the context of the solar corona, in understanding the efficiency of Parker's braiding mechanism for coronal heating.
\item
No exact equilibrium can exist in a periodic domain for the magnetic braids discussed herein. However, approximately force-free states with comparatively small plasma-$\beta$ can exist. These states must contain thin current layers, whose thickness must decrease as the exact equilibrium is approached (Section \ref{approxfff}).
\item
Following the onset of reconnection, a cascade of current sheets forms -- reconnection at which unbraids the magnetic field. {We demonstrated for the first time that this relaxation is approaching a state of decaying turbulence (Figure \ref{spectra}).} The distribution of current sheets in the domain, and eventual overall plasma heating profile, depends on the braiding pattern \citep{wilmotsmith2011b}. 
\item
The turbulence becomes more developed during the relaxation process when either (i) the magnetic braid is more complex (as measured by e.g.~the topological entropy),  or (ii) the plasma resistivity is smaller (section \ref{turbprop}).
\item
The final state of the turbulent relaxation may not be a linear f.f.f.~depending on the topological degree (t.d.) of the magnetic braid {and distribution of field line helicity. The} t.d.~is preserved during the relaxation, and may be inconsistent with the linear force-free state. For the magnetic field of Equation (\ref{beq}), the t.d.~is 2, and the final state consists of a pair of oppositely-twisted flux tubes. {Even when the t.d. does not preclude the Taylor state the field line helicities may.}
\end{enumerate}

\acknowledgments

All the authors acknowledge financial support from the UK's
STFC (grant number ST/K000993).
We are grateful for particularly fruitful discussions with A.~L.~Wilmot-Smith and A.~R.~Yeates.


\end{document}